\documentclass[11pt]{article}
\usepackage{amsmath}
\usepackage{amssymb}
\usepackage{latexsym}
\usepackage{epsfig}

\newcommand{\be}{\begin{equation}}
\newcommand{\ee}{\end{equation}}

\begin{document}
{}~ \hfill\vbox{\hbox{CTP-SCU/2014008}}\break
\vskip 3.0cm
\centerline{\Large \bf Quantum gravity corrections to the tunneling}
\vspace*{1.0ex}
\centerline{\Large \bf radiation of scalar particles}
\vspace*{10.0ex}
\centerline{\large Peng Wang, Haitang Yang and Shuxuan Ying}
\vspace*{7.0ex}
\vspace*{4.0ex}
\centerline{\large \it Center for theoretical physics}
\centerline{\large \it Sichuan University}
\centerline{\large \it Chengdu, 610064, China} \vspace*{1.0ex}
\centerline{pengw@scu.edu.cn, hyanga@scu.edu.cn, 2013222020007@stu.scu.edu.cn}
\vspace*{10.0ex}
\centerline{\bf Abstract} \bigskip \smallskip
The original derivation of Hawking radiation shows the complete evaporation of black holes. However, theories of quantum gravity predict the existence of the minimal observable length. In this paper, we investigate the tunneling radiation of the scalar particles by introducing quantum gravity effects influenced by the generalized uncertainty principle. The Hawking temperatures are not only determined by the properties of the black holes, but also affected by the quantum numbers of the emitted particles. The quantum gravity corrections slow down the increase of the temperatures. The remnants are found during the evaporation.
\vfill
\eject
\baselineskip=16pt
\vspace*{10.0ex}
\tableofcontents

\section{Introduction}

Hawking radiation is described as a quantum tunneling procedure near the horizons of black holes. The original research of the radiation is relied on the calculation of the Bogoliubov coefficient \cite{SWH}. The standard radiation spectrum was derived as the black body spectrum. This result predicts the complete evaporation of the black hole. Other derivations based on the relativity quantum mechanics in curved spacetime or the anomalous in the quantum theory also showed the correctness of the Hawking formula \cite{DR,IUW}. The semi-classical tunneling method is an effective model to study Hawking radiation \cite{KW,PW,ZZ,JWC,KM,KM2,KSAE,LRC}. The varied background spacetime was taken into account in this method, therefore, the corrected Hawking temperature was obtained \cite{PW}. The corrected temperature is higher than the original one, which implies that varied spacetime accelerates the evaporation.

However, various theories of quantum gravity, such as string theory, loop quantum gravity and quantum geometry, predict the existence of the minimal observable length \cite{PKT,ACV,KPP,LJG,GAC}. This view is also supported by Gedanken experiments \cite{S}. The generalized uncertainty principle (GUP) is an simply way to realize this minimal length. To derive GUP, the fundamental commutation relations should be modified. There are two ways to modify the commutation relations, which correspond to different expressions of GUP \cite{KMM,DV}. One way was put forward by Kempf et al. and the fundamental length was derived \cite{KMM}. Based on doubly special relativity theory, Das et al. recently re-modified the relations and derived a expression of GUP, namely, DSR-GUP \cite{DV}. The minimum measurable length and the maximum measurable momentum were approached in the same expression.

Incorporating GUP into black holes, many fruits have been achieved\cite{ACS,BG,BJM,SA,LW,CWY,HN,LLS}. The minimal mass was discussed in the modified Schwarzschild spacetime characterized by a finite Kretschmann scalar in \cite{LLS}. In \cite{HN}, the authors studied the creation of scalar particles pair by an electric field in the presence of a minimal length. The thermodynamical properties were investigated in \cite{BG,LW}. In the previous work \cite{NS}, taking into account effects of quantum gravity influenced by DSR-GUP, Nozari and Saghafi adopted semi-calssical tunneling model to investigate quantum tunneling radiation of a massless particle across the horizon of the Schwarzschild black hole. The modified Hawking temperature was gotten. In \cite{CWY}, based on GUP, the fermions' tunneling radiation of various black holes and their residue mass are considered. The results are same with our solutions.

In this paper, considering effects of quantum gravity influenced by GUP, we modify the Klein-Gordon equation in curved spacetime. Then use the modified equation to discuss the tunneling radiation of scalar particles in a Schwarzschild and a Kerr black holes. Our result shows that the corrected Hawking temperatures depend not only on the parameters of the black holes but also on the quantum numbers (mass, energy and angular momentum) of the emitted particles. Quantum gravity corrections slow down the increase of the Hawking temperatures during the evaporation. This correction leads the remnants during the evaporation.

The rest of this paper is organized as follows. In section 2, we modify the Klein-Gordon equation by the modified fundamental commutation relations. In section 3, using the modified equation, we discuss the tunneling radiation of scalar particles in the Schwarzschild black hole. The tunneling radiation in the Kerr black hole is investigated in section 4. The remnants are derived in section 5. Section 6 is devoted to our conclusion.

\section{Generalized Klein-Gordon equation}

In this section, we discuss the influence of quantum gravity effects on the Klein-Gordon equation. Here we adopt the expression of GUP put forward by Kempf et al., namely \cite{KMM},

\begin{eqnarray}
\Delta x \Delta p \geq \frac{\hbar}{2}\left[1+ \beta (\Delta p)^2\right],
\label{eq1.1}
\end{eqnarray}

\noindent where $\beta = \beta_0 \frac{l^2_p}{\hbar^2}$ is a small value, $\beta_0 <10^{5}$ is a dimensionless parameter \cite{LMS} and $l_p$ is the Planck length. Eq. (\ref{eq1.1}) was derived by the modified the commutation relations $\left[x_{\mu},p_{\nu}\right]= i \hbar \delta_{\mu\nu}\left[1+ \beta p^2\right]$, where $x_{\mu}$ and $p_\mu$ are position and momentum operators defined as follows

\begin{eqnarray}
x_{\mu} & = & x_{0\mu},\nonumber\\
p_{\mu} & = & p_{0\mu}\left(1+\beta p_{0}^{2}\right).
\label{eq1.2}
\end{eqnarray}

\noindent $x_{0\mu}$ and $p_{0\mu}$ satisfy the canonical commutation relations $\left[x_{0\mu},p_{0\nu}\right]= i \hbar \delta_{\mu\nu}$.
The four-dimensional form of the Klein-Gordon equation without an electromagnetic field is given by

\begin{equation}
-P^{\mu}P_{\mu}=m^{2}.
\label{eq2.1}
\end{equation}

\noindent To consider the effects of quantum gravity, we rewrite this equation as

\begin{equation}
-\left(i\hbar\right)^{2}\partial^{t}\partial_{t}=\left(i\hbar\right)^{2}\partial^{i}\partial_{i}+m^{2}.
\label{eq2.2}
\end{equation}

\noindent In the theories of quantum gravity, the generalized expression of energy takes on the form  \cite{WG,NK,HBH}

\begin{equation}
E=E\left(1-\beta E^{2}\right)=E\left[1-\beta\left(P^{2}+m^{2}\right)\right],
\label{eq2.4}
\end{equation}

\noindent where the mass-energy shell condition $E^{2}-P^{2}=m^{2}$ was used.  After inserted the modified operators of momentum into above equation, the generalized Klein-Gordon equation is gotten as

\begin{equation}
-\left(i\hbar\right)^{2}\partial^{t}\partial_{t}\Psi=\left[\left(-i\hbar\right)^{2}\partial^{i}\partial_{i}+m^{2}\right]
\left\{1-2\beta\left[\left(-i\hbar\right)^{2}\partial^{i}\partial_{i}+m^{2}\right]\right\}\Psi.
\label{eq2.5}
\end{equation}

\noindent In the following sections, we will adopt the above equation to discuss the tunneling radiation of scalar particles across the horizons of the Schwarzschild and Kerr black holes.

\section{Particle tunnels from the Schwarzschild black hole}

In this section, we consider a particle tunneling from the Schwarzschild black hole. The effects of quantum gravity are taken into account by the influence of GUP. The metric of the Schwarzschild black hole is given by

\begin{equation}
ds^{2}=-f\left(r\right)dt^{2}+\frac{1}{g\left(r\right)}dr^{2}+r^{2}d\theta^{2}+r^{2}\sin^{2}\theta d\phi^{2},
\label{eq3.1}
\end{equation}

\noindent with $f\left(r\right)=g\left(r\right)=1-\frac{2M}{r}$ and $M$ is the mass of the black hole. The horizon is located at $r_{h}=2M$. The motion of a scalar particle obeys the generalized Klein-Gordon equation (\ref{eq2.5}). To describe the equation of motion, we assume the wave function of the scalar particle as

\begin{equation}
\Psi=\exp\left[\frac{i}{\hbar}I\left(t,r,\theta,\phi\right)\right],
\label{eq3.2}
\end{equation}

\noindent where $I$ is the action of the scalar particle. Inserting the inverse metric into Eq. (\ref{eq2.5}), we can get

\begin{eqnarray}
\frac{1}{f}\left(\partial_{t}I\right)^{2} & = & \left[g\left(\partial_{r}I\right)^{2}+\frac{1}{r^{2}}\left(\partial_{\theta}I\right)^{2}+\frac{1}{r^{2}\sin^{2}\theta}\left(\partial_
{\phi}I\right)^{2}+m^{2}\right] \nonumber \\
 &  & \times\left\{1-2\beta\left[g\left(\partial_{r}I\right)^{2}+\frac{1}{r^{2}}\left(\partial_{\theta}I\right)^{2}+\frac{1}{r^{2}\sin^{2}\theta}
 \left(\partial_{\phi}I\right)^{2}+m^{2}\right]\right\}.
\label{eq3.3}
\end{eqnarray}

\noindent In the above equation, the first order terms of $\hbar$ were neglected. Our aim it to get the expression of the action. It is difficult to solve it from the above equation. To solve this equation, we consider the properties of the space-time and carry out separation of variables on the action $I$ as

\begin{equation}
I=-\omega t+W\left(r\right)+J\left(\theta,\phi\right),
\label{eq3.4}
\end{equation}

\noindent where $\omega$ is the energy of the emitted particle. Here we let the particle tunnel along the radial direction. Thus we can get

\begin{equation}
\frac{1}{r^{2}}\left(\partial_{\theta}J\right)^{2}+\frac{1}{r^{2}\sin^{2}\theta}\left(\partial_{\phi}J\right)^{2}=c.
\label{eq3.5}
\end{equation}

\noindent In the above equation, $c$ is an constant and can be zero. Then Eq. (\ref{eq3.3}) reduces to a quartic equation, that is

\begin{equation}
A\left(\partial_{r}W\right)^{4}+B\left(\partial_{r}W\right)^{2}+C=0,
\label{eq3.6}
\end{equation}

\noindent where

\begin{eqnarray}
A & = & -2\beta g\left(r\right)^{2},\nonumber \\
B & = & g\left(r\right)\left(1-4\beta c-4\beta m^{2}\right),\nonumber \\
C & = & m^{2}+c-2\beta c^{2}-4\beta cm^{2}-2\beta m^{4}-\frac{\omega^{2}}{f\left(r\right)}.
\label{eq3.7}
\end{eqnarray}

\noindent Due to $B^{2}-4AC=1-8\beta\frac{\omega^{2}}{f\left(r\right)}>0$, the condition of solution of the above equation (\ref{eq3.6}) is satisfied. The solution of this quartic equation at the horizon is

\begin{eqnarray}
W_{\pm} & = & \pm\int dr\frac{1}{\sqrt{fg}}\sqrt{\omega^{2}-m^{2}f+2\beta m^{4}f}\left[1+\beta\left(m^{2}+\frac{\omega^{2}}{f}\right)\right]\nonumber \\
& = & \pm i\pi2M\omega\left[1+\frac{1}{2}\beta\left(m^{2}+4\omega^{2}\right)\right]+\left(\mathrm{real\: part}\right),
\label{eq3.8}
\end{eqnarray}

\noindent where $+/-$ represent the outgoing and ingoing solutions, and $f=g=1-\frac{2M}{r}$. Here we adopt the canonically invariant expression of the tunneling rate and let $p_r = \partial_r W$. Thus the tunneling rate is gotten as

\begin{eqnarray}
\Gamma &\propto & exp \left[-Im\oint p_rdr\right]\nonumber\\
&=& exp \left[-Im\left(\int p^{out}_rdr - \int p^{in}_rdr\right)\right]\nonumber\\
&=& exp\left\{-\pi 4M\omega\left[1+\frac{1}{2}\beta\left(m^{2}+4\omega^{2}\right)\right] \right\}.
\label{eq3.9}
\end{eqnarray}

\noindent This is the Boltzmann factor with the Hawking temperature taking $T=\frac{1}{4\pi M\left[1+\frac{1}{2}\beta\left(m^{2}+4\omega^{2}\right)\right]}$. When $\beta = 0$, the result reduces to the usual result. There is two times of the Hawking temperature appeared. However, this result is different from that derived by Hawking. In \cite{APGS}, the authors pointed out that the temporal contribution to the tunneling amplitude was missed in the above discussion. Therefore, we should incorporate this contribution into the calculation of the tunneling rate.

This temporal contribution can be found by using the Kruskal coordinates $(T, R)$. In this coordinates, the exterior region of the black hole is described by

\begin{eqnarray}
T= e^{\kappa r_*}sinh(\kappa t), \quad R= e^{\kappa r_*}cosh(\kappa t),
\label{eq3.10}
\end{eqnarray}

\noindent where $\kappa = 1/4M$ is the surface gravity and $r_* = r +\frac{1}{2\kappa}\ln\frac{r-r_h}{r_h}$ is the tortoise coordinate. Correspondingly, the inner region is given by

\begin{eqnarray}
T= e^{\kappa r_*}cosh(\kappa t), \quad R= e^{\kappa r_*}sinh(\kappa t).
\label{eq3.11}
\end{eqnarray}

\noindent The temporal contribution can be derived by connecting these two patches across the horizon. We rotate the time as $t\rightarrow t - \frac{i\pi} {2 \kappa} $. This rotation would leads to that the temporal part yields an additional imaginary contribution, namely, $Im \omega \Delta t^{out,in} = \frac{\pi \omega } {2\kappa}$. Thus, the total contribution is $Im \omega \Delta t = \pi \omega/\kappa $. Incorporating this contribution into Eq. (\ref{eq3.9}), we get the tunneling rate as

\begin{eqnarray}
\Gamma &\propto & exp \left[-Im\left(\omega t+\oint p_rdr\right)\right]\nonumber\\
&=& exp\left\{-\pi 8M\omega\left[1+\beta\left(\omega^{2}+m^{2}/4\right)\right] \right\}.
\label{eq3.12}
\end{eqnarray}

\noindent This is the Boltzmann factor with the Hawking temperature taking

\begin{eqnarray}
T= \frac{1}{8\pi M\left[1+\beta\left(\omega^{2}+m^{2}/4\right)\right]}=T_0\left[1-\beta\left(\omega^{2}+m^{2}/4\right)\right].
\label{eq3.13}
\end{eqnarray}

\noindent Clearly, the corrected temperature is lower than the original derived one. The corrected temperature is related to the mass and energy of the emitted particle. The quantum correction slows down the increase of the temperature.

\section{Tunneling in the Kerr black hole}

In this section, we investigate the tunneling across the event horizon
of a Kerr black hole with quantum gravity effects. The metric of the Kerr black hole is given by

\begin{eqnarray}
ds^{2} & = & -\left(1-\frac{2Mr}{\rho^{2}}\right)dt^{2}+\frac{\rho^{2}}{\triangle}dr^{2}+\left[\left(r^{2}+a^{2}\right)+\frac{2Mra^{2}\sin^{2}
\theta}{\rho^{2}}\right]\sin^{2}\theta d\varphi^{2}\nonumber \\
 &  & +\rho^{2}d\theta^{2}-\frac{4Mra\sin^{2}\theta}{\rho^{2}}dtd\varphi,
\label{eq4.1}
\end{eqnarray}

\noindent where

\noindent
\begin{eqnarray}
\rho^{2} & = & r^{2}+a^{2}\cos^{2}\theta,\nonumber \\
\triangle & = & r^{2}-2Mr+a^{2}=\left(r-r_{+}\right)\left(r-r_{-}\right).
\label{eq4.2}
\end{eqnarray}

\noindent $r_{\pm}=M\pm\sqrt{M^{2}-a^{2}}$ are the outer and inner horizons, $M$ is the black hole's mass and $a$ is the angular
momentum per unit mass. To calculate the tunneling radiation of a scalar particle, we can directly adopt the metric (\ref{eq4.1}). Here we discuss it in the dragging coordinate coordinate system. Performing the dragging coordinate transformation $\phi=\varphi-\Omega t$, where

\begin{eqnarray}
\Omega=\frac{\left(r^{2}+a^{2}-\triangle\right)a}{\left(r^{2}+a^{2}\right)^{2}-\triangle a^{2}\sin^{2}\theta},
\label{eq4.3}
\end{eqnarray}

\noindent to the metric (\ref{eq4.1}), we get

\begin{eqnarray}
ds^{2} & = & -\frac{\triangle\rho^{2}}{\left(r^{2}+a^{2}\right)^{2}-\triangle a^{2}\sin^{2}\theta}dt^{2}+\frac{\rho^{2}}{\triangle}dr^{2}+\rho^{2}d\theta^{2}+\frac{\left(r^{2}+a^{2}\right)^{2}-\triangle a^{2}\sin^{2}\theta}{\rho^{2}}\sin^{2}\theta d\phi^{2},\nonumber \\
 & \equiv & -F\left(r\right)dt^{2}+\frac{1}{G\left(r\right)}dr^{2}+K\left(r\right)^2d\theta^{2}+H\left(r\right)^2d\phi^{2}.
\label{eq4.4}
\end{eqnarray}

\noindent We also assume the wave function of scalar field takes a form

\begin{eqnarray}
\psi=\exp\left[\frac{i}{\hbar}I\left(t,r,\theta,\phi\right)\right],
\label{eq4.5}
\end{eqnarray}

\noindent where $I$ is the scalar particle's action. We take the inverse metric and (\ref{eq4.5}) into (\ref{eq2.5}), and ignore the higher order of $\hbar$. Then we get

\begin{eqnarray}
\frac{1}{F}\left(\partial_{t}I\right)^{2} & = & \left[G\left(\partial_{r}I\right)^{2}+\frac{1}{K{}^{2}}\left(\partial_{\theta}I\right)^{2}+\frac{1}{H{}^{2}}\left(\partial_{\phi}I\right)^{2}
+m^{2}\right]\times\nonumber \\
 &  & \left(1-2\beta\left[G\left(\partial_{r}I\right)^{2}+\frac{1}{K{}^{2}}\left(\partial_{\theta}I\right)^{2}+\frac{1}{H{}^{2}}\left(\partial_{\phi}I\right)^{2}
 +m^{2}\right]\right).
\label{eq4.6}
\end{eqnarray}

\noindent To solve the above equation, we carry out the separation of variables as follows

\begin{eqnarray}
I=-\left(\omega-j\Omega\right)t+W\left(r,\theta\right)+j\phi,
\label{eq4.7}
\end{eqnarray}

\noindent where $\omega$ is the energy and $j$ is the angular momentum of the emitted particle. It is worth to note that $W\left(r,\theta\right)$
can not be separated as $W\left(r\right)\Theta\left(\theta\right)$. We fix the angle $\theta$ at a certain value of $\theta_{0}$. Therefore, our aim is to find the solution of $W\left(r,\theta_{0}\right)$. Substituting the action (\ref{eq4.7}) into Eq. (\ref{eq4.6}), we get

\begin{eqnarray}
A\left(\partial_{r}W\right)^{4}+B\left(\partial_{r}W\right)^{2}+C=0,
\label{eq4.8}
\end{eqnarray}

\noindent where

\begin{eqnarray}
A & = & -2\beta G^{2},\nonumber \\
B & = & G\left(1-4\beta\frac{j^{2}}{H^{2}}-4\beta m^{2}\right),\nonumber \\
C & = & m^{2}+\frac{j^{2}}{H^{2}}-2\beta\left(\frac{j^{2}}{H^{2}}\right)^{2}-4\beta m^{2}\frac{j^{2}}{H^{2}}-2\beta m^{4}-\frac{\left(\omega-j\Omega\right)^{2}}{F}.
\label{eq4.9}
\end{eqnarray}

\noindent Solving the above equation at event horizon yields

\begin{eqnarray}
W_{\pm} & = & \pm\int dr\frac{1}{\sqrt{FG}}\sqrt{\left(\omega-j\Omega\right)^{2}-m^{2}F-\frac{j^{2}}{H{}^{2}}F\left(r\right)+2\beta\left(F\frac{j^{4}}
{H{}^{4}}+2m^{2}F\frac{j^{2}}{H{}^{2}}+m^{4}F\left(r\right)\right)}\nonumber \\
 &  & \times\left[1+\beta\left(m^{2}+\frac{\left(\omega-j\Omega\right)^{2}}{F}+\beta\frac{j^{2}}{H{}^{2}}\right)\right]\nonumber \\
 & = & \pm i\pi\left(\omega-j\Omega_{+}\right)\frac{r_{+}^{2}+a^{2}}{r_{+}-r_{-}}\left(1+\beta\Pi\right)+\left(\mathrm{real\: part}\right),
\label{eq4.10}
\end{eqnarray}

\noindent where

\begin{eqnarray}
\Pi & = & \frac{1}{2\left(r_{+}^{2}+a^{2}\right)^{5}\left(r_{+}-r_{-}\right)^{2}\omega_{0}^{4}\left(r_{+}^{2}+a^{2}\cos^{2}\theta_{0}\right)^{2}}\left\{ \left(r_{+}-r_{-}\right)\left(aj-\left(r_{+}^{2}+a^{2}\right)\omega\right)^{4}\left(r_{+}^{2}+a^{2}\cos^{2}\theta_{0}\right)\times\right.\nonumber \\
 &  & \left[\left(r_{+}^{2}+a^{2}\right)\left(r_{+}-r_{-}\right)\left(r_{+}^{2}+a^{2}\cos^{2}\theta_{0}\right)\left(m^{2}+\frac{j^{2}\left(r_{+}^{2}+a^{2}
 \cos^{2}\theta_{0}\right)^{2}}{\left(r_{+}^{2}+a^{2}\right)^{4}}\right)+\right.\nonumber \\
 &  & \left.+2aj\omega_{0}\left(-\left(r_{+}^{2}+a^{2}\right)\left(3r_{+}-r_{-}\right)+a^{2}\left(r_{+}-r_{-}\right)\sin^{2}\theta_{0}\right)\right]\nonumber \\
 &  & \omega_{0}^{2}\left(r_{+}^{2}+a^{2}\right)\left(\left(r_{+}^{2}+a^{2}\right)\omega-aj\right)^{3}\left(\left[-8aj\left(r_{+}-r_{-}\right)
 \left(r_{+}^{2}+a^{2}\cos^{2}\theta_{0}\right)\right]\times\right.\nonumber \\
 &  & \left[-\left(r_{+}^{2}+a^{2}\right)\left(3r_{+}-r_{-}\right)+a^{2}\left(r_{+}-r_{-}\right)\sin^{2}\theta_{0}\right]-\left(\left(r_{+}^{2}+a^{2}\right)
 \omega-aj\right)\times\nonumber \\
 &  & \left[a^{2}\cos^{2}\theta_{0}\left(4\left(r_{+}^{2}+a^{2}\right)\left(a^{2}-2r_{+}^{2}+3r_{+}r_{-}\right)+3a^{2}\left(r_{+}-r_{-}\right)^{2}\sin^{2}\theta_{0}
 \right)+\right.\nonumber \\
 &  & \left.\left.\left.r_{+}\left(4\left(r_{+}^{2}+a^{2}\right)\left(-r_{+}^{2}\left(r_{+}-2r_{-}\right)+a^{2}\left(2r_{+}-r_{-}\right)\right)+3a^{2}r_{+}
 \left(r_{+}-r_{-}\right)^{2}\sin^{2}\theta_{0}\right)\right]\right)\right\} ,
\label{eq4.11}
\end{eqnarray}

\noindent $\Omega_{+}=\frac{a}{r_{+}^{2}+a^{2}}$ is the angular velocity at the outer horizon and $\omega_{0}=\omega-j\Omega_{+}$. Here, we still use the canonically invariant expression of the tunneling rate. To take into account the temporal contribution, we adopt the Kruskal coordinates $(T,R)$. The exterior region $(r>r_+)$ of the Kerr black hole is described by the Kruskal coordinates as follows

\begin{eqnarray}
T &=& e^{\kappa_+r_*}sinh(\kappa_+t),\nonumber\\
R &=& e^{\kappa_+r_*}cosh(\kappa_+t),
\label{eq4.12}
\end{eqnarray}

\noindent where $r_*= r+ \frac{1}{2\kappa_+}ln\frac{r-r_+}{r_+}- \frac{1}{2\kappa_-}ln\frac{r-r_-}{r_-}$ is the tortoise coordinate, and $\kappa_\pm =\frac{r_+ - r_-}{2(r_{\pm}^2+a^2)}$ is the surface gravity at the outer (inner) horizons. The interior region is described as

\begin{eqnarray}
T &=& e^{\kappa_+r_*}cosh(\kappa_+t),\nonumber\\
R &=& e^{\kappa_+r_*}sinh(\kappa_+t).
\label{eq4.13}
\end{eqnarray}

\noindent To find the temporal contribution, we connect the above two patches across the horizon. We rotate the time as $t\rightarrow t - \frac{i\pi} {2 \kappa_+} $. This rotation leads to the additional imaginary contribution, $Im E \Delta t^{out,in} = \frac{\pi E } {2\kappa_+}$, where $E=\omega - j\Omega_+$. Therefore, the total contribution is $Im E \Delta t = \frac{\pi E } {\kappa_+}$. Incorporating this contribution into Eq. (\ref{eq3.9}), we get the tunneling rate as

\begin{eqnarray}
\Gamma &\propto & exp \left[-Im\left(E t+\oint p_rdr\right)\right]\nonumber\\
& = & \exp\left[-4\pi\left(\omega-j\Omega_{+}\right)\frac{r_{+}^{2}+a^{2}}{r_{+}-r_{-}}\left(1+\beta\Pi/2\right)\right].
\label{eq4.14}
\end{eqnarray}

\noindent This is the Boltzmann factor with the Hawking temperature taking

\begin{equation}
T= \frac{r_{+}-r_{-}}
{4\pi\left(r_{+}^{2}+a^{2}\right)\left(1+\beta\Pi/2\right)}=\left(1-\beta\Pi/2\right)T_{0},
\label{eq4.15}
\end{equation}

\noindent where $T_{0}=\frac{\left(r_{+}-r_{-}\right)}{4\pi\left(r_{+}^{2}+a^{2}\right)}$ is the original Hawking temperature of the Kerr black hole. It is clearly that the corrected temperature is lower than the original one. The corrected temperature is not only dependent one the quantum number (energy, mass, angular momentum) of the emitted fermion, but also related to the azimuthal angle $\theta$.

\section{Remnants in black holes}

In this section, we derive the remnants in the black holes. The quantum correction slows down the increase of the Hawking temperature due to the radiation. Finally, there is a balance state. At this state, the evaporation stops and remnants are produced. When $a=0$, the Hawking temperature of the Kerr black hole reduce to that of the Schwarzschild black hole.

\begin{equation}
T=\frac{1}{8\pi M}\left[1-\beta\left(\omega^{2}+m^{2}/4\right)\right],
\label{eq5.1}
\end{equation}

\noindent To determine the remnants, it is enough to adopt the massless particle. The evaporation stops when the following condition

\begin{equation}
\left(M-dM\right)\left(1+\beta\omega^{2}\right)\simeq M,
\label{eq5.2}
\end{equation}

\noindent is satisfied. To avoid the appearance of the negative temperature, $\omega \leq 1/\sqrt{\beta}$. Then with the observation $\beta = \beta_0/M_p$ and $dM = \omega$, where $M_p$ is the Planck mass and $\beta_0<10^5$ is a dimensionless parameter, we get the remnants and the temperature

\begin{equation}
M_{\mathrm{Res}}\simeq\frac{M_{p}^{2}}{\beta_{0}\omega}\gtrsim\frac{M_{p}}{\sqrt{\beta_{0}}}.
\label{eq5.3}
\end{equation}

\noindent Thus, the singularity of the black hole is prevented by the quantum gravity correction. Due to the value of $\beta_0$, the temperature of the remnants may be higher than the Planck temperature.

\section{Conclusion}
In this paper, we modified the generalized Klein-Gordon equation in curved spacetime based on the generalized uncertainty principle \cite{KMM}. Then used the modified equation, we investigated the tunneling radiation of the scalar particles in the Schwarzschild and Kerr black holes. In these two black hole spacetime configurations, we showed that the corrected Hawking temperatures does not only depend on the properties of the black holes, but also depend on the quantum numbers (angular momentum, mass, energy) of the emitted particles. Our results show that the temperature increasing during the evaporation will be slowed down by the quantum gravity effects. At a certain point, the temperature becomes balanced and remnants of the black holes exist. The remnants were derived as $M_{\hbox{Res}}\gtrsim\frac{M_p}{\sqrt{\beta_0}}$ by the emission of the massless particle. In \cite{CA}, the remnants were also derived by different methods.

\bigskip
\noindent
{\bf Acknowledgements}
This work is supported in part by NSFC (Grant No. 11005016, 11175039 and 11375121) and SYSTF (Grant No. 2012JQ0039).


\begin{thebibliography}{99}


\bibitem{SWH}
S.W. Hawking, \emph{Particle creation by black holes}, \emph{Commun. Math. Phys.} \textbf{43} (1975) 199.

\bibitem{DR}
T. Damoar and R. R uffini, \emph{Phys. Rev.} \textbf{D 14} (1976) 332.

\bibitem{IUW}
S.P. Robinson and F. Wilczek, \emph{Relationship between Hawking radiation and gravitational anomalies}, \emph{Phys. Rev. Lett.} \textbf{95} (2005) 011303.

S. Iso, H. Umetsu and F. Wilczek, \emph{Hawking radiation from charged black holes via gauge and gravitational anomalies}, \emph{Phys. Rev.} {\bf D 74}  (2006) 044017.

\bibitem{KW}
P.~Kraus and F.~Wilczek, \emph{Effect of selfinteraction on charged black hole radiance}, \emph{Nucl. Phys.} {\bf B 437} (1995) 231.

\bibitem{PW}
M.K.~Parikh and F.~Wilczek, \emph{Hawking radiation as tunneling}, \emph{Phys. Rev. Lett. } {\bf 85} (2000) 5042.

\bibitem{ZZ}
 J.~Zhang and Z.~Zhao, \emph{Charged particles' tunnelling from the Kerr-Newman black hole}, \emph{Phys. Lett.} {\bf B 638} (2006) 110.

\bibitem{JWC}
Q.Q.~Jiang, S.Q.~Wu and X.~Cai, \emph{Hawking radiation as tunneling from the Kerr and Kerr-Newman black holes}, \emph{Phys. Rev. } {\bf D 73} (2006) 064003.

\bibitem{KM}
R.~Kerner and R.B.~Mann, \emph{Fermions tunnelling from black holes}, \emph{Class. Quant. Grav.} {\bf 25} (2008) 095014.

\bibitem{KM2}
R.~Kerner and R.B.~Mann, \emph{Tunnelling, temperature and Taub-NUT black holes}, \emph{Phys. Rev.} {\bf D 73} (2006) 104010.

\bibitem{KSAE}
S.~Hemming and E.~Keski-Vakkuri, \emph{The spectrum of strings on BTZ black holes and spectral flow in the SL(2,R) WZW model}, \emph{Nucl. Phys.} {\bf B 626} (2002) 363.

A.J.M.~Medved, \emph{Radiation via tunneling from a de Sitter cosmological horizon}, \emph{Phys. Rev.} {\bf D 66} (2002) 124009.

E.C.~Vagenas, \emph{Semiclassical corrections to the Bekenstein-Hawking entropy of the BTZ black hole via selfgravitation}, \emph{Phys. Lett.} {\bf B 533} (2002) 302.

M.~Arzano, A.J.M.~Medved and E.C.~Vagenas, \emph{Hawking radiation as tunneling through the quantum horizon}, \emph{JHEP} {\bf 0509} (2005) 037.

S.Q.~Wu and Q.Q.~Jiang, \emph{Remarks on Hawking radiation as tunneling from the BTZ black holes}, \emph{JHEP} {\bf 0603} (2006) 079.

M.~Nadalini, L.~Vanzo and S.~Zerbini, \emph{Hawking radiation as tunneling: The D dimensional rotating case}, \emph{J. Phys.} {\bf A 39} (2006) 6601.

B.~Chatterjee, A.~Ghosh and P.~Mitra, \emph{Tunnelling from black holes in the Hamilton-Jacobi approach}, \emph{Phys. Lett.} {\bf B 661} (2008) 307.

V. Akhmedova,  T. Pilling, A. de Gill and D. Singleton, \emph{Comments on anomaly versus WKB/tunneling methods for calculating Unruh radiation}, \emph{Phys. Lett. } \textbf{B 673} (2009) 227.

R.~Banerjee and B.R.~Majhi, \emph{Quantum tunneling and back reaction}, \textit{ Phys. Lett.} {\bf B 662} (2008) 62.

D.~Singleton, E.C.~Vagenas, T.~Zhu and J.R.~Ren, \emph{Insights and possible resolution to the information loss paradox via the tunneling picture},
 \emph{JHEP} {\bf 1008} (2010) 089.

\bibitem{LRC}
R.~Li and J.R.~Ren, \emph{Dirac particles tunneling from BTZ black hole}, \emph{Phys. Lett.}  {\bf B 661} (2008) 370.

D.Y.~Chen, Q.Q.~Jiang and X.T.~Zu, \emph{Hawking radiation of Dirac particles via tunnelling from rotating black holes in de Sitter spaces},
 \emph{Phys. Lett.} {\bf B 665} (2008) 106.

D.Y.~Chen, H.~Yang and X.T.~Zu, \emph{Hawking radiation of black holes in the z = 4 Horava-Lifshitz gravity}, \emph{Phys. Lett.} {\bf B 681} (2009) 463.

Q.Q.~Jiang, \emph{Dirac particles' tunnelling from black rings}, \emph{Phys. Rev.}  {\bf D 78} (2008) 044009.

K.~Lin and S.Z.~Yang, \emph{Fermion tunneling from higher-dimensional black holes}, \emph{Phys. Rev.} {\bf D 79} (2009) 064035.

R.~Di Criscienzo and L.~Vanzo, \emph{Fermion tunneling from dynamical horizons}, \emph{Europhys. Lett.}  {\bf 82} (2008) 60001.

S.A.~Hayward, R.D. Criscienzo, L.~Vanzo et al, \emph{Local Hawking temperature for dynamical black holes}, \emph{ Class\ Quant\ Grav.}  {\bf 26} (2009) 062001.

\bibitem{PKT}
P.K. Townsend, \emph{Small-scale structure of spacetime as the origin of the gravitational constant}, \emph{Phys. Rev.} \textbf{D 15} (1977) 2795.

\bibitem{ACV}
D. Amati, M. Ciafaloni and G. Veneziano, \emph{Can spacetime be probed below the string size?}, \emph{Phys. Lett.}  \textbf{B 216} (1989) 41.

\bibitem{KPP}
K. Konishi, G. Paffuti and P. Provero, \emph{Minimum physical length and the generalized uncertainty principle in string theory}, \emph{Phys. Lett.}  \textbf{B 234} (1990) 276.

\bibitem{LJG}
L.J.~Garay, \emph{Quantum gravity and minimum length}, \emph{Int. J. Mod. Phys.} {\bf A 10} (1995) 145.

\bibitem{GAC}
G.~Amelino-Camelia, \emph{Relativity in space-times with short distance structure governed by an observer independent (Planckian) length scale},
\emph{Int. J. Mod. Phys.} {\bf D 11} (2002) 35.

\bibitem{S}
F. Scardigli, \emph{Phys. Lett.} {\bf B 452} (1999) 39.

\bibitem{KMM}
A.~Kempf, G.~Mangano and R.B.~Mann, \emph{Hilbert space representation of the minimal length uncertainty relation}, \emph{ Phys. Rev.} {\bf D 52} (1995) 1108.

\bibitem{DV}
S. Das, E. C. Vagenas, \emph{Phys. Rev. Lett.} {\bf 101}  (2008) 221301 [arXiv:0810.5333 [hep-th]].

\bibitem{ADV}
A. F. Ali, S.Das and E. C. Vagenas, \emph{Phys. Lett.} {\bf B 678} (2009) 497.

\bibitem{ACS}
R.J.~Adler, P.~Chen and D.I.~Santiago, \emph{The generalized uncertainty principle and black hole remnants}, \emph{Gen. Rel. Grav.}  \textbf{33} (2001)  2101.

\bibitem{BG}
R.~Banerjee and S.~Ghosh, \emph{Generalised Uncertainty Principle, remnant mass and singularity problem in black hole thermodynamics}, \emph{Phys. Lett.}
{\bf B 688} (2010) 224.

\bibitem{BJM}
A.~Bina, S.~Jalalzadeh and A.~Moslehi, \emph{Quantum black hole in the Generalized Uncertainty Principle framework}, \emph{Phys. Rev.} {\bf D 81} (2010) 023528.

\bibitem{SA}
J.L.~Said and K.Z.~Adami, \emph{The Generalized Uncertainty Principle in f(R) gravity for a charged black hole}, \emph{Phys. Rev.} {\bf D 83} (2011) 043008.

\bibitem{LW}
L.~Xiang and X.Q.~Wen, \emph{Black hole thermodynamics with generalized uncertainty principle}, \emph{JHEP} {\bf 0910} (2009) 046.

\bibitem{NS}
K.~Nozari and S.~Saghafi, \emph{Natural cutoffs and quantum tunneling from black hole horizon}, \emph{JHEP} {\bf 1211} (2012) 005.

\bibitem{CWY}
  D.~Chen, H.~Wu and H.~Yang,
  ``Fermion's tunnelling with effects of quantum gravity,''
  Adv.\ High Energy Phys.\  {\bf 2013}, 432412 (2013)
  [arXiv:1305.7104 [gr-qc]].\\
  D.~Chen, H.~Wu and H.~Yang,
  ``Observing remnants by fermions' tunneling,''
  JCAP {\bf 1403}, 036 (2014)
  [arXiv:1307.0172 [gr-qc]].\\
  D.~Chen,
  ``Dirac particles` tunneling from five-dimensional rotating black strings influenced by the generalized uncertainty principle,''
  Eur.\ Phys.\ J.\ C {\bf 74}, 2687 (2014)
  [arXiv:1312.2075 [hep-th]].\\
  D.~Y.~Chen, Q.~Q.~Jiang, P.~Wang and H.~Yang,
  ``Remnants, fermions` tunnelling and effects of quantum gravity,''
  JHEP {\bf 1311}, 176 (2013)
  [arXiv:1312.3781 [hep-th]].\\
  D.~Chen and Z.~Li,
  ``Remarks on Remnants by Fermions' Tunnelling from Black Strings,''
  Adv.\ High Energy Phys.\  {\bf 2014}, 620157 (2014)
  [arXiv:1404.6375 [hep-th]].



\bibitem{HN}
S. Haouat and K. Nouicer, \emph{Influence of a Minimal Length on the Creation of Scalar Particles}, \emph{Phys. Rev.} \textbf{D 89} (2014) 105030 [arXiv:1310.6966[hep-th]].

\bibitem{LLS}
X. Li, Y. Ling and Y.G. Shen, \emph{Singularities and the Finale of Black Hole Evaporation}, \emph{Int. J. Mod. Phys.} \textbf{D 22} (2013), 1342016, [ arXiv:1305.3851[gr-qc]].

\bibitem{LMS}
S.~Liberati, L.~Maccione and T.P.~Sotiriou, \emph{Scale hierarchy in Horava-Lifshitz gravity: a strong constraint from synchrotron
radiation in the Crab nebula}, \emph{Phys. Rev. Lett.} {\bf 109} (2012) 151602.

\bibitem{WG}
W. Greiner, Relativistic Quantum Mechanics: Wave Equation, Springer-Verlag, 2000.

\bibitem{NK}
K.~Nozari and M.~Karami, \emph{Minimal length and generalized Dirac equation}, \emph{Mod. Phys. Lett.} {\bf A 20} (2005) 3095.

\bibitem{HBH}
S.~Hossenfelder, M.~Bleicher, S.~Hofmann, et. al., \emph{Collider signatures in the Planck regime}, \emph{Phys. Lett.} {\bf B 575} (2003) 85.

\bibitem{CA}
P.~Chen and R.J.~Adler, \emph{Black hole remnants and dark matter}, \emph{Nucl. Phys. Proc. Suppl.}  {\bf 124} (2003) 103.

L.~Xiang, \emph{A note on the black hole remnant}, \emph{Phys. Lett.}  \textbf{B 647} (2007) 207.



\bibitem{APGS}
V. Akhmedova, T. Pilling, A. de Gill and D. Singleton, \emph{Temporal contribution to gravitational WKB-like calculations}, \emph{Phys. Lett.} \textbf{B 666} (2008) 269.

E.T. Akhmedov, T. Pilling and D. Singleton, \emph{Subtleties in the quasi-classical calculation of Hawking radiation}, \emph{Int. J. Mod. Phys.} \textbf{D 17} (2008) 2453.






\end{thebibliography}
\end{document}